\documentclass{optica-article}

\journal{opticajournal}

\articletype{Research Article}

\usepackage{lineno}

\begin{document}

\title{Observation of Q-switched and continuous wave regimes with mode-hopping in Er-doped fiber lasers incorporating a dynamic population grating}

\author{Zengrun Wen,\authormark{1,2,4} Xiulin Fan,\authormark{1,2,4} Kaile Wang,\authormark{3} Weiming Wang\authormark{1,2} Song Gao,\authormark{1,2} Wenjing Hao,\authormark{1} Yuanmei Gao\authormark{1,2}, Yangjian Cai,\authormark{1,2,5} and Liren Zheng\authormark{1,6}}

\address{\authormark{1}Shandong Provincial Engineering and Technical Center of Light Manipulation \& Shandong Provincial Key Laboratory of Optics and Photonic Devices, School of Physics and Electronics, Shandong Normal University, Jinan 250358, China\\
\authormark{2}Collaborative Innovation Center of Light Manipulation and Applications, Shandong Normal University, Jinan 250358, China\\
\authormark{3}The State Key Laboratory of Integrated Service Networks, The School of Telecommunications Engineering, Xidian University, Xian, 710071, China\\
\authormark{4}The authors contribute equally to this work.}
\email{\authormark{5}yangjiancai@sdnu.edu.cn\\
\authormark{6}zlrgym@sdnu.edu.cn}

\begin{abstract}
Dynamic population gratings (DPGs) in rare-earth doped fibers are prevalent devices in fiber lasers for the production of single-longitudinal-mode emission, Q-switched pulses, and wavelength self-sweeping regimes. This study presents a transition from Q-switched state to continuous wave (CW) state, accompanying irregular mode-hopping, in an erbium-doped fiber laser with a heavily-doped DPG centered at 1549.95 nm. Our results demonstrate that the transition between these two states can be achieved by adjusting the pump power. The repetition frequency of the Q-switched pulse increases monotonically with the increasing pump power, while the pulse duration initially narrows and then expands because the reduced peak intensity weakens the nonlinear effect. Additionally, modulation peaks are evident on both the Q-switched pulse train and the CW background, which are induced by the irregular mode-hopping caused by the DPG. Furthermore, we observe that the central wavelength fluctuates within a range of 0.05 nm. These results provide valuable insight into the DPG effect in heavily-doped fibers.
\end{abstract}

\section{Introduction}
Narrow linewidth lasers are widely utilized in a variety of applications, including optical communication, coherent beam combination, coherent Doppler LIDAR, sensing, spectroscopy, high-resolution measurement, and so on~\cite{al-taiyUltranarrow2014,heHighPowerCoherent2006,yang100mWLinear2012,weldon1997,ouslimaniNarrow2015,sharmaFiber2005}. Among the different types of lasers, fiber lasers have a great advantage in achieving narrow linewidth emission, due to the variety of filters that can be designed to restrict the number of lasing modes flexibly. Some of the most commonly used filters for fundamental linewidth compression are narrow fiber Bragg gratings and bandpass filters~\cite{kramerFemtosecond2019}. Other filters, such as the Lyot filter~\cite{jingImpedance2017,jiangWidely2022}, multimode interference filter~\cite{mohammedAllfiber2006,kubotaTunable2021}, and Fabry-Perot filter~\cite{mongaStable2021}, can also be used to further compress the lasing wavelength or even obtain single-longitudinal-mode (SLM) operation through tunable periodic filtering. As an alternative design, a compound ring cavity structure incorporating several cascaded or nested fiber rings to form narrow filters~\cite{fengFourwavelength2019,fengWavelength2021}. In addition, rare-earth doped fibers can also be employed to narrow the linewidth through the ASE gain compression effect and dynamic population grating (DPG)~\cite{yaoNarrowbandwidth,stepanov2011}.

Among the above filters, DPGs have been widely studied as a promising tool for narrowing the linewidth in fiber lasers. DPGs, which are formed in rare-earth doped fibers due to standing wave fields, take the advantage of high compatibility with optical fiber systems and flexibility in wavelength tuning~\cite{poozeshHighSNR2018,wangWidelyTunable2020}. Correspondingly, there has been numerous discussion around the mechanism of DPGs~\cite{fanBandwidth2005,stepanov2004,xuTransient2012,lobachOpencavity2017,poozeshSinglef2019}. Recent studies showed that there are two types of amplitude DPGs, namely absorption and gain types, that provide wavelength stabilization and unstable frequency hops, respectively~\cite{zhangSingle2022a,poddubrovskiiRegular2022}. On the other hand, phase DPGs break the symmetry of the optical spectrum and shift the reflective peak, resulting in regular mode-hopping~\cite{drobyshevSpectral2019}. Whereas the wavelength stabilization effect has facilitated the development of fiber laser configurations that utilize DPGs, such as single-frequency fiber lasers and wavelength-tunable lasers~\cite{wangSinglefrequency2021,wangWavelength2019}. Once the DPG is induced in gain fibers, the regular mode-hopping occurs so as to achieve wavelength self-sweeping (WLSS) fiber lasers with microsecond self-pulsation in temporal domain~\cite{kashirinaDual2020a}. In addition, DPG in heavily doped fiber enables the pulsed regime due to the nonradiative transition induced by active-ion clusters~\cite{kurkovEffect2010,tsaiAllfiber2009,tsaiSelfQswitched2009}. Such DPG makes it possible to achieve Q-switched or mode-locked pulses with narrow linewidth~\cite{wenNarrowBandwidth2021}.

In order to stabilize the fiber laser incorporating a dynamic population grating (DPG) as the saturable absorber, it is necessary to separate the gain fiber and DPG to avoid the formation of gain DPG~\cite{poddubrovskii2022}. To achieve the goal, a sigma-cavity setup, which deploys a unidirectional ring for the gain fiber and a linear part for the DPG, isolated by a circulator, is commonly employed in stable single-frequency fiber lasers~\cite{li212kHzlinewidth2018,zhangSingle2022a}. However, the sigma-cavity has not been entirely successful in preventing the formation of gain DPG, which arises from the standing wave resulting from the interference of forward light and faint reflective feedback from the linear part~\cite{peterkaSelfinduced2012}. As a result, wavelength self-sweeping (WLSS) operations have been observed instead of stable single frequency emission in such configurations~\cite{wenSelfsweeping2021}. This raises the question of determining the exact boundaries between stable lasing and WLSS operation in different fiber lasers with sigma-cavity. In addition, mode-hopping in Er-doped fiber is challenging to achieve as the phase DPG in Er-doped fiber (EDF) is very weak~\cite{poddubrovskii2022}. Despite this difficulty, researchers have still achieved CW WLSS operation and a transition from single longitudinal mode (SLM) operation to regular mode-hopping dynamics in Er-doped fiber lasers with emission near 1560 nm~\cite{kashirinaDual2020a,poddubrovskiiRegular2022,poddubrovskiiTimeresolved2022}. The study reveals that the SLM regime is closely related to the high dopant concentration, fiber length of the DPG, and pump power. When a heavily doped fiber is used as the DPG in a sigma-cavity, the situation becomes more complex due to the possibility of pulsed regimes. However, to date, the mode-hopping behavior has not been reported in Er-doped fiber lasers with DPG induced in heavily doped fibers.

In this study, we report a transition from Q-switch to continuous wave (CW) regimes with irregular mode-hoppings in an Er-doped fiber laser with a sigma-cavity configuration where a heavily doped DPG is incorporated. The spectral width of the DPG is limited to 0.27 nm and is centered at 1549.5 nm. By adjusting the polarization controller (PC) and increasing the pump power, the laser transitions from the Q-switching regime to the CW state. The central wavelength variation is measured to be within 0.05 nm, which results in changes in the pulse dynamics of the Q-switched pulse train and the CW background. Further, the radio frequency spectra and optical spectra are analyzed to verify the laser regimes.

\section{Experimental setup}
\begin{figure}[h!]
\centering\includegraphics[width=12cm]{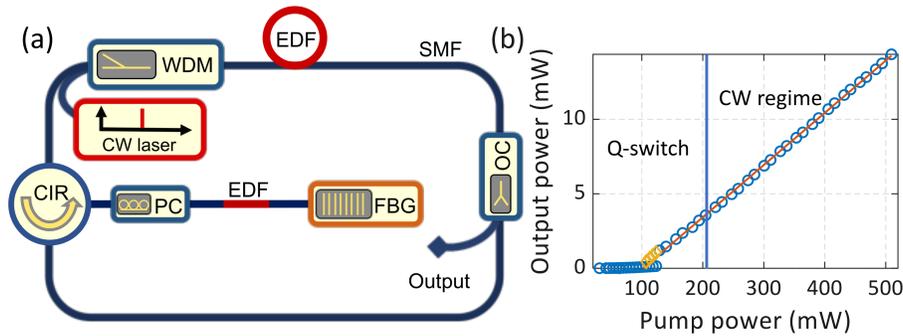}
\caption{Experimental setup and average output power of the Er-doped fiber laser. (a) Schematic diagram pf the sigma-cavity containing a ring part and a linear part. CW: continuous wave; WDM: wavelength division multiplex; EDF: erbium-doped fiber; SMF: single-mode fiber; OC: optical coupler; CIR: circulator; PC: polarization controller; FBG: fiber Bragg grating. (b) Average power versus pump power.}
\label{setup}
\end{figure}
In the present experiment, we employ a standard sigma-cavity configuration for the Er-doped fiber laser, as depicted in Fig.\ref{setup}(a). The cavity comprises a fiber ring and a linear part. The fiber ring contains a 0.6 m-long erbium-doped fiber (EDF, Er110-4/125) that is pumped by a CW laser diode at the wavelength of 975 nm, via a 980/1550 nm wavelength division multiplexing (WDM) device. The laser diode provides a continuously tunable pump power from 0 to 509.4 mW. An optical coupler (OC) splits 30\% of the light energy for the output, while the remaining 70\% circulates within the cavity. The ring cavity is connected to the linear part via a circulator (CIR), which enables unidirectional light propagation. In the linear part, the polarization direction is controlled by using a PC. A 0.18 m-long EDF (Er110-4/125) serves as both the saturable absorber and the DPG. Finally, the laser output is further reflected by a fiber Bragg grating (FBG) with a central wavelength of 1549.95 nm and a full-width half-maximum bandwidth of 0.27 nm. All components of the cavity are well spliced by single-mode fibers (SMF, SMF-28e). Unlike previous relevant studies that were performed near the wavelength of 1560 nm, the present experiment restricts the lasing spectrum to the vicinity of 1550 nm. This difference seriously affects the absorption and reflection properties of DPG. Additionally, using a heavily doped fiber for the saturable absorber enables the formation of Q-switched regimes~\cite{wenNarrowBandwidth2021}. The SMF in the linear part of the cavity is $\sim$5 m longer than in the previous architecture, enabling the control of oscillating modes~\cite{stepanov2011}. A PC in the cavity allows for manipulating lasing polarization states, differing from our previous work~\cite{wenGenerating2021}. The average power, temporal pulse train, optical spectra, and radio frequency signals are measured using a power meter (Thorlabs, PM100D) with a photodiode sensor (Thorlabs, S148C), a real-time digital storage oscilloscope (DSO, Teraxion), an optical spectrum analyzer (OSA, YOKOGAWA, AQ6370D), and a frequency spectrum analyzer (FSA, Agilent, N9320B), respectively. 

As depicted in Fig.~\ref{setup}(b), the average output power of the laser is measured as increasing the pump power. The laser can self-start at a pump power of 129.0 mW and initially operates in a Q-switched mode with mode-hopping. Further gradually increasing the pump power to 220.7 mW, the laser operates at CW regime with mode-hopping instead of Q-switched pulses. There are two bistable states during the transition process between different regimes. Upon reduction of the pump power, a transition from the CW to the Q-switched regime is observed at a pump power of 215.3 mW. Additionally, the laser ceases to emit any output when the pump power is reduced to 107.2 mW. A linear relationship between the output power and pump power is observed, with a slope of 3.5\%. The low slope can be attributed to the high loss induced by the narrow reflect linewidth and high absorption of the heavily doped EDF.

\section{Results and Discussion}
\begin{figure}[h!]
	\centering\includegraphics[width=13cm]{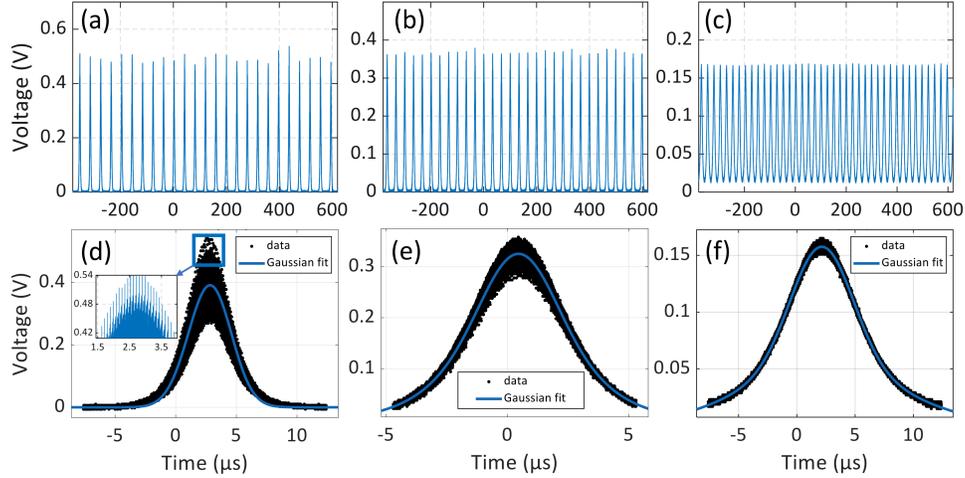}
	\caption{Temporal pulse performances of Q-switched regimes. (a)-(c) Pulse trains and (d-f) the corresponding single pulse profiles at pump power of 183.0 mW, 193.8 mW and 209.9 mW, respectively.}
	\label{q-switch-osc}
\end{figure}
In Fig.\ref{q-switch-osc}, three typical results of the Q-switched performance are presented by gradually modulating the pump power from 129.0 mW to 220.7 mW. Figure~\ref{q-switch-osc}(a) exhibits a Q-switched pulse train with obvious peak fluctuation, which manifests an unstable Q-switch state when the pump power is 183.0 mW. This indicates that the laser intensity is insufficient to sustain a stable DPG. The single pulse duration is 1.81 $\upmu$s and strong noise signals are presented, as illustrated in Fig.~\ref{q-switch-osc}(d). Additionally, there is a series of peaks with an interval equivalent to the cavity period located on the Q-switched pulse envelope (as shown in the inset of Fig.~\ref{q-switch-osc}(d)). When the pump power is raised to 193.8 mW, the repeat frequency of the pulse train increases from 25.5 kHz to 30.0 kHz and the pulse duration shortens from 1.81 $\upmu$s to 1.49 $\upmu$s, which are typical characteristics of passive Q-switching [see in Figs.~\ref{q-switch-osc}(b) and (e)]. If the pump power is further increased, the repeat frequency of the pulse train continues to increase, and the peak intensity decreases, as displayed in Fig.~\ref{q-switch-osc}(c), resulting in a more stable pulse train. However, the pulse duration broadens to 2.62 $\upmu$s, and the background of the Q-switched pulses is above zero ($\sim$0.015 V). The broadened pulse duration is caused by the weakened nonlinearity that originates from the reduction of peak intensity, indicating that the laser simultaneously emits a Q-switched pulse and CW lasing.

\begin{figure}[h!]
	\centering\includegraphics[width=10cm]{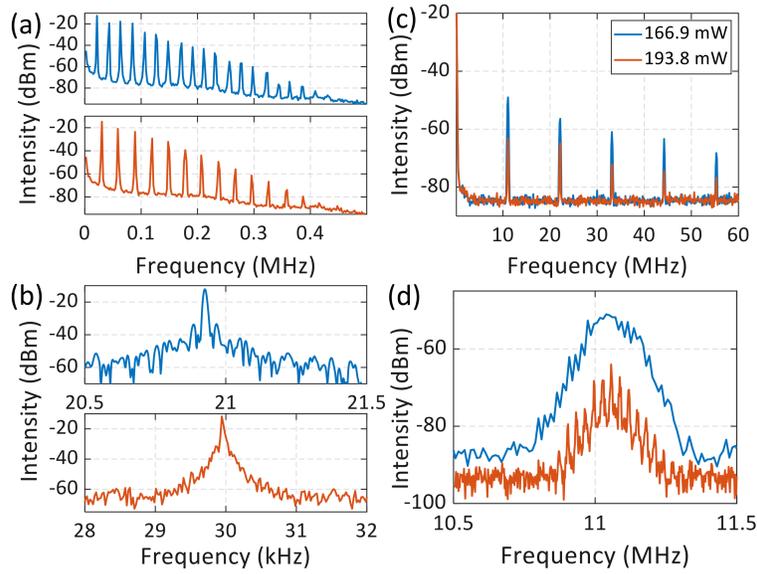}
	\caption{RF spectra of the Q-switched regimes with mode-hopping. (a) RF signals; (b) fundamental frequencies; (c) cavity mode sequences and (d) fundamental cavity modes of the Q-switched pulses.}
	\label{q-switch-esa}
\end{figure}
To validate the results of the Q-switched regime, the radio frequency (RF) spectra were measured using an FSA, as shown in Fig.~\ref{q-switch-esa}. In Fig.~\ref{q-switch-esa}(a), the typical frequency spectra of the Q-switch at a pump power of 166.9 mW and 193.8 mW are displayed. The high-order frequency signals expand in the frequency range of 0-0.4 MHz, and the corresponding fundamental frequencies of 44.2 kHz and 49.3 kHz are shown in Fig.~\ref{q-switch-esa}(b). The signal-to-noise ratios (SNRs) of these two Q-switched states are above 21 dB at a resolution bandwidth of 10 Hz, indicating the unstable of the laser signals. The cavity mode frequencies at different pump powers are displayed in Fig.~\ref{q-switch-esa}(c), where it shows that, for various pump powers, the mode frequencies remain constant, only with an obvious reduction in intensity. This phenomenon corresponds to the decreasing peaks on the Q-switched envelope as the pump power is increased. In Fig.~\ref{q-switch-esa}(d), the fundamental mode frequencies are displayed, and many peaks are observed, which is a result of multiple reflections caused by the DPG.

\begin{figure}[h!]
	\centering\includegraphics[width=10cm]{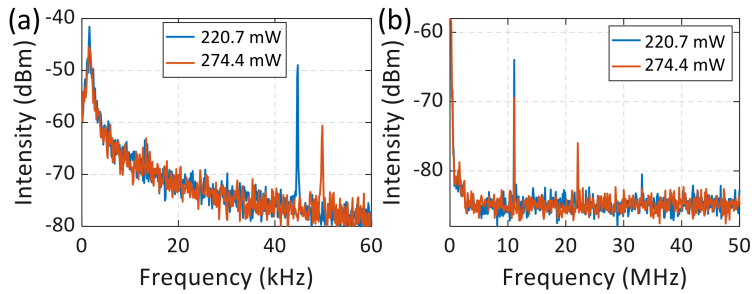}
	\caption{RF spectra of the CW regimes with mode-hopping. (a) RF signals within 60 kHz. (b) Cavity mode sequences.}
	\label{cw-esa}
\end{figure}
As the pump power is further increased, the laser transitions from a Q-switched pulse train to a CW regime. The RF spectra at pump powers of 220.7 mW and 274.4 mW are displayed in Fig.~\ref{cw-esa}, which demonstrate the disappearance of the Q-switched frequency sequence, being replaced by only two isolated, weak pulsating signals located at approximately 42 kHz and 49 kHz, respectively. This phenomenon indicates the breakdown of the Q-switched pulse sequence. Furthermore, the mode frequency in the CW regime is further reduced, as depicted in Fig.~\ref{cw-esa}(b). 

\begin{figure}[h!]
	\centering\includegraphics[width=13cm]{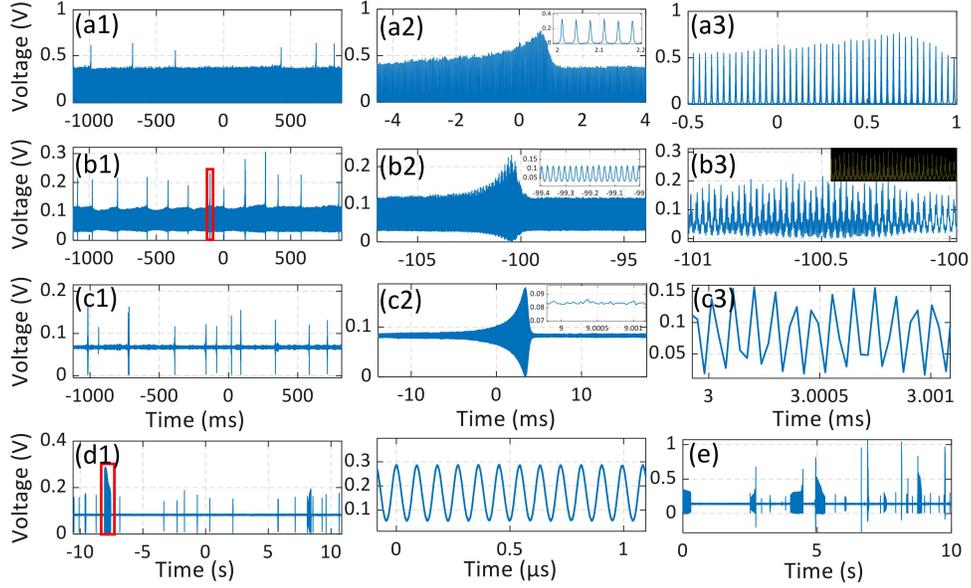}
	\caption{Pulse intensity dynamics at different pump powers. Pulse dynamics in long time scale (a1, b1, c1), single mode-hopping pulse envelope (a2, b2, c2) and the corresponding magnifications (a3, b3, c3) at pump power of 193.8 mW (top row), 215.3 mW (second row) and 220.7 mW (third row), respectively. (d1) Pulse dynamics and (d2) its local detail at pump power of 274.4 mW. (e) Pulse dynamics at pump power of 353.4 mW. The insets in (a2, b2, c2) are the local details in non-mode-hopping areas.}
	\label{cw-osc}
\end{figure}
The pulse dynamics of the Q-switched and CW regimes with mode-hopping are illustrated in Figure~\ref{cw-osc}. Compared to ytterbium-doped fiber lasers, the frequency of mode-hopping in the laser is relatively low (several Hertz), whereas it increases as the pump power is increased [as seen in Figs.~\ref{cw-osc}(a1), (b1), and (c1)]. Furthermore, the mode-hopping time (duration of the pulse envelope) is several milliseconds, which is much larger than that observed in other rare-earth doped fiber lasers, as demonstrated in Figs.~\ref{cw-osc}(a2)-(c2). The local enlarged details of Figs.~\ref{cw-osc}(a2)-(c2) are displayed in Figs.~\ref{cw-osc}(a3)-(c3), respectively. In the mode-hopping areas, the pulse dynamics also alter obviously against the pump power. At relatively low pump powers [193.8 mW], mode-hopping mainly results in the intensity enhancement of the Q-switched pulses. With increasing pump power, in addition to the intensity enhancement, new pulses emerge near the Q-switched pulses, causing intensity modulation [as shown in Figs.~\ref{cw-osc}(b3)]. In both cases, the Q-switched pulses in the non-mode-hopping area remain stable, as shown in the insets of Figs.~\ref{cw-osc}(a2) and (b2). In Figs.~\ref{cw-osc}(c2) and (c3), the non-mode-hopping area is the CW output, while an interference signal with a frequency consistent with the cavity frequency is present at mode-hopping area, which is a typical characteristic of mode-hopping or WLSS fiber lasers~\cite{poddubrovskiiRegular2022}. When the pump power is increased above 274.4 mW, broad interference signals [indicated by the red box in Fig.~\ref{cw-osc}(d1)] begin to emerge, as seen in Figs.~\ref{cw-osc}(d1)-(e). The oscillation frequency of the broad interference signals is consistent with the cavity length. This type of signal can be stable in all polarization-maintaining configurations~\cite{wangSinglefrequency2021}, but instability here is caused by polarization change. As a result, the mode-hopping frequency becomes less stable and decreases.

\begin{figure}[h!]
	\centering\includegraphics[width=10cm]{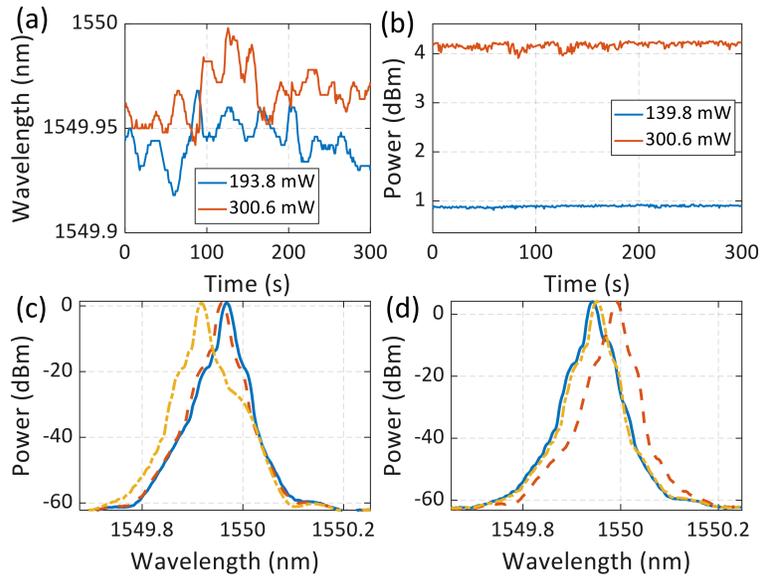}
	\caption{Optical spectra at pump power of 193.8 mW and 300.6 mW, respectively. (a) Variations of central wavelengths within 300 s. (b) Powers stability within 300 s. (c) and (d) the retrieved spectral profiles at the two pump powers.}
	\label{osa-dyn}
\end{figure}
The optical spectra of the laser are evaluated by measuring the central wavelength and its power over a period of 300 minutes, as depicted in Figs.~\ref{osa-dyn}(a) and (b). The results showed an irregular variation of the optical spectrum, indicating that the emission is not in the WLSS regime. The range of wavelength variation varied with changes in pump power, indicating the presence of implicit birefringent filtering in the cavity. The central wavelength was found to become unstable as the pump power was increased, as can be seen in Fig.~\ref{osa-dyn}(b). The optical spectra at different times for pump powers of 193.8 mW and 300.6 mW are displayed in Figs.~\ref{osa-dyn}(c) and (d), respectively, and it can be observed that there is an obvious change in the central wavelength, accompanied by slight variations in the waveform. The 3dB spectrum width remained almost constant with a value of about 0.035 nm, which is consistent with the calculated reflective spectrum width of the DPG based on the absorption-emission cross-section and the Kramers-Kronig relation~\cite{wenGenerating2021}. This result suggests that the DPG indeed plays a crucial role in filtering.

\section{Conclusion}
We induce a DPG in a piece of heavily-doped erbium-doped fiber (EDF), and observe the transition in a widely used sigma-cavity fiber laser. The pump power is utilized as a control parameter to drive the laser from the Q-switched regime to the CW regime, which is accompanied by irregular mode-hopping. As the pump power was increased, the repetition frequency and peak intensity of the Q-switched regime decreased monotonically, while the pulse duration first narrowed and then broadened. At the same time, the peaks on the Q-switched envelopes are diminished, and a CW background begin to emerge. The mode-hopping results in an increase in intensity and the emergence of new pulses based on the Q-switched pulse train. Upon the breakdown of the Q-switched state, the CW regime with mode-hopping is generated, and its pulse dynamics were consistent with those observed in regular mode-hopping or WLSS fiber lasers. The central wavelength was limited to near 1550 nm by the fiber Bragg grating and varied within 0.05 nm, along with a slight variation in the spectral structure. This study provides an attempt on experimentally studying DPGs in heavily doped fibers. The intrinsic mechanisms of DPGs in heavily doped fibers is more complexity due to the existence of ion cluster effects for Q-switched regimes, which need further study.

\begin{backmatter}
\bmsection{Funding}
We acknowledge the support of Key Research and Development Program of China (2022YFA1404 800, 2019YFA0705000); National Natural Science Foundation of China (12104272, 12274270, 91950104, 12192254, 92250304, 11974218); Local science and technology development project of the central government (YDZX20203700001766).

\bmsection{Acknowledgments} We appreciate professor Zuoqiang Hao and Zhanghua Han for the support of measurement devices.

\bmsection{Disclosures}
The authors declare no conflicts of interest.

\bmsection{Data Availability Statement}
Data underlying the results presented in this paper are not publicly available at this time but may be obtained from the authors upon reasonable request.

\end{backmatter}


\bibliography{refs}






\end{document}